\documentclass[11pt]{article}

\usepackage{graphicx}
\usepackage{amsmath}
\usepackage{amssymb}
\usepackage{dcolumn}
\usepackage{bm}
\usepackage{slashed}
\usepackage{authblk}
\usepackage{hyperref}

\newcommand{\be}{\begin{equation}}
\newcommand{\ee}{\end{equation}}
\newcommand{\ba}{\begin{eqnarray}}
\newcommand{\ea}{\end{eqnarray}}
\newcommand{\no}{\nonumber \\}
\newcommand{\gsim}{\mathrel{\hbox{\rlap{\lower.55ex \hbox {$\sim$}}
                   \kern-.3em \raise.4ex \hbox{$>$}}}}
\newcommand{\lsim}{\mathrel{\hbox{\rlap{\lower.55ex \hbox {$\sim$}}
                   \kern-.3em \raise.4ex \hbox{$<$}}}}

\def\roughly#1{\mathrel{\raise.3ex\hbox{$#1$\kern-.75em%
\lower1ex\hbox{$\sim$}}}}
\def\lsim{\roughly<}
\def\gsim{\roughly>}

\def\({\left(}
\def\){\right)}
\def\[{\left[}
\def\]{\right]}
\def\<{\langle}
\def\>{\rangle}

\def\h{{\eta}}

\def\l{{\lambda}}
\def\L{{\Lambda}}
\def\d{{\delta}}
\def\D{{\Delta}}
\def\o{{\omega}}
\def\O{{\Omega}}
\def\e{{\epsilon}}

\def\a{{\alpha}}
\def\b{{\beta}}
\def\c{{\chi}}

\def\g{{\gamma}}
\def\G{{\Gamma}}
\def\p{{\pi}}

\def\m{{\mu}}
\def\n{{\nu}}
\def\r{{\rho}}
\def\s{{\sigma}}

\def\t{{\tau}}
\def\th{{\theta}}

\def\ps{{\psi}}

\newcommand{\pd}{{\partial}}

\newcommand{\feq}{f_\text{eq}}
\newcommand{\cV}{{\cal V}}
\newcommand{\cA}{{\cal A}}

\date{\today}

\begin{document}

\title{\bf Transient spin modes from relaxational axial kinetic theory}

\author[1]{Shu Lin\thanks{linshu8@mail.sysu.edu.cn}}
\author[1]{Haiqin Tang}
\affil[1]{School of Physics and Astronomy, Sun Yat-Sen University, Zhuhai 519082, China}

\maketitle

\begin{abstract}
We study the dynamics of spin mode by solving the axial kinetic equations under the relaxation time approximation in the presence of dissipative sources. We find transient spin modes in response to electric field with spacetime inhomogeneity, fluid acceleration and shear. To the lowest order in spatial momentum $k$, we find the responses to electric field and acceleration can be interpreted as retarded response to temporal variations of magnetic field and vorticity respectively. The response to shear occurs at $O(k^2)$ and can be reduced to retarded response to spatial variation of vorticity. Beyond lowest order, the responses to all three sources are non-local with branch cut in the dispersions. We argue that the non-locality is a consequence of the quasi-particle picture underlying the kinetic description. We also analyze spin transport equation taking into account spin response to temporal and spatial variations of vorticity. We find the corrections turn the original first order spin transport equation into a third order one (or a second order one in the homogeneous limit). The change in order of transport equation is a consequence of non-local nature of the responses, suggesting possible breakdown of gradient expansion in spin hydrodynamics for microscopic theories with quasi-particles. 
\end{abstract}


\newpage

\section{Introduction}

It has been pointed out that the final particles should be spin polarized owing to spin-orbital coupling in off-central heavy ion collisions \cite{Liang:2004ph,Liang:2004xn}. Indeed, this has been confirmed in global spin polarization of $\L$-hyperon in relativistic heavy ion collisions (HIC) \cite{STAR:2017ckg}, triggering intensive theoretical studies in the polarization phenomenon. The global polarization data have been well understood with the spin-vorticity coupling picture, witnessing the quark-gluon plasma (QGP) as a most vortical fluid \cite{Becattini:2013fla, Fang:2016vpj, Li:2017slc, Liu:2019krs, Becattini:2017gcx, Wei:2018zfb, Wu:2019eyi, Fu:2020oxj, Zhang:2019xya, Weickgenannt:2020aaf}. However, the same mechanism fails to account for the local spin polarization of $\L$-hyperon measured in HIC experiments \cite{STAR:2019erd}, with phenomenological studies predicting a pattern with the wrong sign \cite{Becattini:2017gcx,Wei:2018zfb,Fu:2020oxj}. More recent studies reveal a more complete picture: spin has momentum dependent couplings to almost all hydrodynamic gradient \cite{Hidaka:2017auj,Liu:2021uhn,Becattini:2021suc,Weickgenannt:2022zxs}, which does not affect the momentum-averaged global polarization, but does contribute to the momentum-dependent local polarization. In particular, the spin-shear coupling gives the correct sign of local polarization and may lead to the correct pattern when combined with spin-vorticity coupling \cite{Fu:2021pok,Becattini:2021iol,Yi:2021ryh,Fu:2022myl,Wu:2022mkr}.

Current phenomenological studies are based on the spin coupling to shear in static and homogeneous limit, which have been obtained in field theory \cite{Liu:2021uhn,Becattini:2021suc} and quantum kinetic theory \cite{Hidaka:2017auj}, with the implicit assumption that the spin density relaxes in a much shorter time scale than the hydrodynamic scale. The opposite limit of slow spin relaxation has also been considered: the limit allows one to include spin as extra slow degrees of freedom in conventional hydrodynamics \cite{Florkowski:2017ruc,Hattori:2019lfp,Fukushima:2020ucl,Hongo:2021ona,Biswas:2023qsw,She:2021lhe,Cao:2022aku}, with thermal vorticity treated as the corresponding spin potential \cite{Becattini:2013fla,Becattini:2018duy,Li:2020eon}. The slow relaxation limit has indeed be confirmed in a microscopic computation for heavy quark \cite{Hongo:2022izs}. It is of both theoretical and phenomenological interests to study the situation in between the two limits. On one hand, in the existing spin hydrodynamic framework, spin couples to vorticity only\footnote{The statement depends on definition of spin tensor. It holds for total anti-symmetric definition of spin tensor, which we also take in this work.}. This is a consequence of homogeneous limit, in which the spin coupling to other hydrodynamic gradients for single particle vanishes after the isotropic phase space integration. Beyond the homogeneous limit, the spin coupling to other hydrodynamic gradients are generically allowed. On the other hand, being only approximately conserved at most, the spin tensor is expected to have a dispersion different from strictly conserved hydrodynamic modes. In particular, in a microscopic theory with quasi-particles, we expect branch cut to be present in the dispersion. The aim of this work is to illustrate these two aspects by deriving explicit transient spin modes using axial kinetic theory (AKT) \cite{Hattori:2019ahi,Weickgenannt:2019dks,Gao:2019znl,Yang:2020hri}.

The ideal spin hydrodynamics have been derived using collisionless axial kinetic theory in \cite{Peng:2021ago}. Dissipative spin hydrodynamics has also been derived using collisional kinetic theory and classical description of spin \cite{Bhadury:2020cop,Weickgenannt:2022qvh,Weickgenannt:2022zxs,Bhadury:2022ulr}, see also dissipative spin hydrodynamics from chiral kinetic theory \cite{Shi:2020htn}. We will instead focus on the regime where spin mode is faster than the hydrodynamic modes. We will also consider electric field, which bears close similarity with the hydrodynamic sources. We will consider a probe fermion subject to above mentioned sources and use relaxation time approximation (RTA) for collision term in the AKT to simplify the analysis.

The paper is organized as follows: In Section~\hyperref[sec_rta]{2}, we give a brief review of the axial kinetic theory used in this work and specify the RTA approximation for the collision term for electromagnetic and hydrodynamic sources respectively. In Section~\hyperref[sec_transient]{3}, we solve the axial kinetic equations and perform phase space integration to obtain the transient spin modes. Physical interpretation of these modes will be discussed. In Section~\hyperref[sec_spinhydro]{4}, we discuss implications of the transient spin modes for spin hydrodynamics. We summarize in Section~\hyperref[sec_sum]{5}. Some detailed calculations are collected in appendix.

\section{Axial kinetic theory with relaxation time approximation}\label{sec_rta}

The collisionless AKT has been derived from Wigner function  defined as
\begin{align}
S^<_{\a\b}(p,X)=\int d^4Y e^{ip\cdot Y}\<\bar{\ps}_\b(y)U(y,x)\ps_\a(x)\>,
\end{align}
where $\ps$ is Dirac spinor for fermions. $U(y,x)=\exp\(-i\int_x^y dz^\r A_\r(z)\)$ is the gauge link connecting two fermion fields at coordinates $x$ and $y$. $X=\frac{x+y}{2}$ and $Y=x-y$ correspond to average and difference of the two coordinates. The Wigner function can be decomposed in the Clifford algebra as
\begin{align}
S^<={\cal S}+{\cal P}i\g^5+\cV_\m\g^\m+\cA_\m\g^5\g^\m+\frac{{\cal S}_{\m\n}}{2}\s^{\m\n}.
\end{align}
Not all components are independent. A common choice for degrees of freedom is the vector and axial components $\cV^\m$ and $\cA^\m$. Their momentum integrations give rise to vector and axial current as
\begin{align}
J^\m=4\int_p\cV^\m,\quad J_5^\m=4\int_p\cA^\m.
\end{align}
with $\int_p=\int\frac{d^4p}{(2\p)^4}$. $J_5^\m$ can be related to canonical definition of spin tensor as
\begin{align}
S^{\l\m\n}=\frac{1}{8}\<\bar{\ps}\{\g^\l,\s^{\m\n}\}\ps\>=\frac{1}{2}\e^{\s\l\m\n}J_{5,\s}.
\end{align}
The spin tensor density in the framework of spin hydrodynamics is defined from part of the spin tensor comoving with the fluid \cite{Hattori:2019lfp}
\begin{align}
S^{\m\n}=S^{\l\m\n}u_\l.
\end{align}
In fluid's rest frame $u^\l=(1,0,0,0)$, $S^{\m\n}$ and $J_5^\m$ are related by $S^{ij}=\frac{1}{2}\e^{ijk}J_{5,k}$. Below we shall refer to $J_5$ as spin density.

The kinetic equations have been derived by several groups \cite{Hattori:2019ahi,Weickgenannt:2019dks,Gao:2019znl,Yang:2020hri}. We will follow the notations of \cite{Hattori:2019ahi}, in which the kinetic equations read
\begin{align}
&\D\cdot \cV=0,\label{eq1}\\
&(p^2-m^2)\cV_\m=-\tilde{F}_{\m\n}\cA^\n,\label{eq2}\\
&p_\n \cV_\m-p_\m \cV_\n=\frac{1}{2}\e_{\m\n\r\s}\D^\r \cA^\s,\label{eq3}\\
&p\cdot \cA=0,\label{eq4}\\
&(p^2-m^2)\cA^\m=\frac{1}{2}\e^{\m\n\r\s}p_\s\D_\n \cV_\r,\label{eq5}\\
&p\cdot\D \cA^\m+F^{\n\m}\cA_\n=\frac{1}{2}\e^{\m\n\r\s}\pd_\s F_{\l\n}\pd_p^\l \cV_\r,\label{eq6}
\end{align}
with $\D_\m=\pd_\m+F_{\n\m}\pd_p^\n$. A distinct feature not present in chiral kinetic theory is that spin becomes an independent degree of freedom in AKT, which is essential for our study.
In equilibrium state without electromagnetic field, we have\footnote{We have suppressed an overall factor of $2\p$, which will be canceled in energy integration $\int\frac{dp_0}{2\p}$ to be performed later.}
\begin{align}
\cV^\m=p^\m\d(p^2-m^2)\feq,\quad \cA^\m=0,
\end{align}
where $\feq=\frac{1}{e^{(p_0-\m)/T}+1}$ being the Fermi-Dirac distribution. $T$ and $\m$ are the temperature and chemical potential respectively. Out of equilibrium, $\cA^\m$ can be induced by electromagnetic or hydrodynamic sources, corresponding to external field effect and off-equilibrium effect respectively. To discuss dissipative sources we need to introduce collision term, which will be done separately for the two sources.

We first consider the electromagnetic source in collisionless AKT. The spin density induced by electromagnetic fields are expected to be $\cA^\m\sim O(F_{\m\n})$. To simplify the analysis, we adopt the power counting $F_{\m\n}\sim O(\pd)$ and restrict up to linear order in $F$. This leads to a decoupled set equations for $V^\m$ from \eqref{eq1} through \eqref{eq3}
\begin{align}\label{Veqs}
&\D\cdot \cV=0,\no
&(p^2-m^2)\cV_\m=0,\no
&p_\n \cV_\m-p_\m \cV_\n=0,
\end{align}
which can be solved by
\begin{align}\label{V_sol}
\cV^\m= p^\m\d(p^2-m^2)f_V,
\end{align}
with $f_V=\feq+O(F)$. In fact the $O(F)$ correction enters \eqref{eq5} and \eqref{eq6} only as quadratic terms in $F$, which we ignore. Thus we can simply use equilibrium solution for $\cV$ on the right hand side (RHS) of \eqref{eq5} and \eqref{eq6}. The constraint equations \eqref{eq5} determine a non-dynamical part of the solution
\begin{align}\label{off-shell}
\cA^\m=\frac{1}{2}\e^{\m\n\r\s}p_\n F_{\r\s}\d'(p^2-m^2)\feq.
\end{align}
This is a trivial generalization of the counterpart in chiral kinetic theory, giving the spin magnetic coupling and spin Hall effect as
\begin{align}
\cA^i=p_0 \(B_i-\e^{ijk}p_jE_k\)\d'(p^2-m^2)\feq,
\end{align}
with $E_i=F_{0i}$ and $B_i=-\frac{1}{2}\e^{ijk}F_{jk}$ being electric and magnetic fields. Note that the remaining dynamical part of the solution can only be determined by solving the transport equation \eqref{eq6} with given initial conditions.
The situation is qualitatively similar in collisional axial kinetic equations \cite{Yang:2020hri}. In the presence of electric field, we still split the solution into a non-dynamical part and dynamical part as
\begin{align}\label{split}
\cA^\m=\d\cA^\m+\frac{1}{2}\e^{\m\n\r\s}p_\n F_{\r\s}\d'(p^2-m^2)f_V.
\end{align}
The second term on the RHS is non-dynamical as it relaxes with the vector density, while the first term $\d\cA^\m$ evolves according to the axial kinetic equation \eqref{eq6}. Following \cite{Bhadury:2020puc}, we will add an RTA type collision term to \eqref{eq6} as
\begin{align}
p\cdot\D \cA^\m+F^{\n\m}\cA_\n=\frac{1}{2}\e^{\m\n\r\s}\pd_\s F_{\l\n}\pd_p^\l \cV_\r-\frac{p\cdot u}{\t}\d\cA^\m,
\end{align}
with $\t$ being the relaxation time.
The form of the relaxation term implies that $\cA^\m$ relaxes to \eqref{off-shell} in the end. This is the expected limit as supported by field theory calculations of spin Hall effect \cite{Liu:2020dxg}\footnote{To compare with Eq.~(20) in \cite{Liu:2020dxg}, one needs to integrate \eqref{off-shell} over $p_0$. Using $\int dp_0 \d'(p^2-m^2)\feq=\int dp_0\frac{\pd}{2p_0\pd p_0}\d(p^2-m^2)\feq=-\int dp_0\(\frac{1}{2p_0^2}+\frac{\pd}{2p_0\pd p_0}\)\feq$, we easily find agreement with long time limit $q_0\to 0$ of Eq.~(20) in \cite{Liu:2020dxg}.}. We will not worry about modification of the constraint equations \eqref{eq4} and \eqref{eq5} with the logic that the constraint equations will remain satisfied once satisfied by the initial conditions. Using \eqref{split}, we easily obtain the dynamical equation for $\d{\cal A}^\m$
\begin{align}\label{transport_RTA}
p\cdot\D \d\cA^\m+\frac{p\cdot u}{\t}\d\cA^\m=\frac{1}{2}\e^{\m\n\r\s}\pd_\s F_{\l\n}\pd_p^\l \cV_\r.
\end{align}

Next we turn to the hydrodynamic source. In this case, the constraint equations \eqref{eq4} and \eqref{eq5} can be solved by \cite{Liu:2021uhn}\footnote{The field theory and kinetic theory approaches differ in the massive case. Given that the present result from the latter approach lacks a contribution from dynamics of spin degree of freedom, we take the result from field theory approach.}
\begin{align}\label{leq_sol}
\cA^\m=\d\(p^2-m^2\)\frac{\e^{\m\n\r\s}p_\a u_\b}{2p\cdot u}\pd_\n f_V,
\end{align}
with $f_V=\frac{1}{e^{p\cdot u/T}+1}$ being local equilibrium distribution. In the absence of collision term, the transport equation \eqref{eq6} is satisfied in non-dissipative vorticity source only. For dissipative sources such as shear and acceleration etc, collision term is needed to balance the transport equation. \eqref{leq_sol} for dissipative sources gives rise to local spin polarization \cite{Lin:2022tma,Lin:2024zik}\footnote{Despite these sources produce entropy in hydrodynamic evolution, spin response to them \eqref{leq_sol} is kinematic involving no entropy production.}. \eqref{leq_sol} should be understood as spin reached in static and homogeneous limit. Beyond this limit, we expect the spin to trace the slowly evolving hydrodynamic sources. Following \cite{Bhadury:2020puc} for the case of vorticity, we will use the following RTA type collision term for the transport equation
\begin{align}\label{transportH_RTA}
p\cdot\pd \cA^\m=-\frac{p\cdot u}{\t}\(\cA^\m-H^\m\),
\end{align}
with the hydrodynamic sources $H^\m$ given by \cite{Hidaka:2017auj,Yi:2021ryh,Liu:2021uhn,Becattini:2021suc}
\begin{align}\label{H_sources}
&H_{\text{shear}}^\m=\d(p^2-m^2)\frac{1}{2p\cdot u}\e^{\m\n\a\b}p_\a u_\b p^\s\s_{\s\n}f_V',\no
&H_{\text{acc}}^\m=-\d(p^2-m^2)\frac{1}{4}\e^{\m\n\a\b}p_\a u_\b Du_\n f_V',\no
&H_\text{temp}^\m=\d(p^2-m^2)\e^{\m\n\a\b}p_\a u_\b\pd_\n \(\frac{1}{T}\)\(\frac{\m}{p\cdot u}\) f_V',\no
&H_{\text{chem}}^\m=-\d(p^2-m^2)\frac{1}{2p\cdot u}\e^{\m\n\a\b}p_\a u_\b\pd_\n\m f_V',
\end{align}
corresponding to shear, acceleration, temperature gradient and chemical potential gradient respectively\footnote{Instead of working with dimensionless thermal vorticity, thermal shear etc, we have shuffled all temperature gradient effect into $H_\text{temp}$}. $f_V'\equiv \frac{\pd}{\pd (p\cdot u)}f_V$. The shear tensor and acceleration are defined by
\begin{align}\label{shear_def}
&\s^{\m\n}=\frac{1}{2}\(\D^{\m\a}\D^{\n\b}+\D^{\m\b}\D^{\n\a}\)\pd_\a u_\b-\frac{1}{3}\D^{\m\n}\D^{\a\b}\pd_{\a}u_\b,\no
&D=u\cdot\pd, 
\end{align}
with $\D^{\m\n}=-g^{\m\n}+u^\m u^\n$ being the projector. 
In the next section, we will solve \eqref{transport_RTA} and \eqref{transportH_RTA}. The relaxation term governs the evolution of spin density towards its long time limit. We will refer to the corresponding spin modes as transient modes below. Furthermore, we will consider inhomogeneous sources, which is crucial for spin density in dissipative hydrodynamic sources and branch cut in dispersion of spin modes.

\section{Transient spin modes}\label{sec_transient}

Now we solve \eqref{transport_RTA} and \eqref{transportH_RTA} for the dynamical part of ${\cal A}^\m$. The resulting solution corresponds to a transient spin mode at specific momentum. We then perform momentum integration to obtain a collective transient spin modes. We first consider \eqref{transport_RTA} and choose the fluid's rest frame. The relevant dissipative source is electric field with spacetime inhomogeneity. The RHS of \eqref{transport_RTA} has been evaluated in the appendix with the following results
\begin{align}\label{source_E}
&s^0=-p_0p_j\e^{ijk}\pd_k F_{0i}\d'(p^2-m^2)f_V+\frac{1}{2}\e^{ijk}\pd_k F_{0i}(-p_j)\d(p^2-m^2)f_V'\no
&s^i=p_ip_j\e^{jkl}\pd_l F_{0k}\d'(p^2-m^2)f_V+m^2\e^{ilk}\pd_k F_{0l}\d'(p^2-m^2)f_V+\frac{1}{2}\e^{ijk}\pd_k F_{0j}p_0\d(p^2-m^2)f_V'\no
&+\frac{1}{2}\e^{ijk}\pd_0 F_{0j}p_k\d(p^2-m^2)f_V'.
\end{align}
%
With the source \eqref{source_E}, \eqref{transport_RTA} can be solved using Fourier transform. Note that $p\gg\pd_X$ means the separation of scale between $p$ and the momentum conjugate to $X$ denoted by $k$: $p\gg k$. We consider response of $\d\widetilde{\cal A}^\m$ to electric field in plane wave form, for which we easily obtain
\begin{align}\label{S_schematic}
\d\widetilde{{\cal A}}^\m=\frac{s^\m}{p_0/\t+i {\bf k}\cdot{\bf p}-i\o p_0},
\end{align}
with $\d\widetilde{\cal A}^\m$ being the Fourier transform of $\d{\cal A}^\m$ with respect to $X$. Its momentum integration gives the Fourier transform of the spin density as
\begin{align}
\tilde{J}_5^\m=4\int_p\d\tilde{\cal A}^\m.
\end{align}
Without loss of generality, we point the electric field in $y$ direction with inhomogeneity along $z$.\footnote{There could be inhomogeneity along $y$, but it has no effect according to \eqref{source_E}.} Note that $s^\m$ contains both $\d(p^2-m^2)$ and $\d'(p^2-m^2)$. The former is used to performed the $p_0$ integral straightforwardly. The latter is treated as
\begin{align}
&\int dp_0 G(p_0)\d'(p^2-m^2)=\int dp_0 G(p_0)\frac{\pd}{2p_0\pd p_0}\d(p^2-m^2)\no
&=-\int dp_0\frac{\pd}{\pd p_0}\(\frac{G(p_0)}{2p_0}\)\d(p^2-m^2),
\end{align}
with $G(p_0)$ includes all $p_0$ dependencies in both numerator and denominator of \eqref{S_schematic} apart from $\d'(p^2-m^2)$. Writing $\int_p=\int \frac{dp_0 p^2dp d\O}{(2\p)^4}$ and performing the angular integration, we find the component $\m=0$ vanishes by rotational invariance, which implies no axial charge production in this case. The component $\m=i$ corresponds to spin modes. The resulting expression is lengthy and not shown here.

\begin{figure}[htbp]
	\begin{center}
		\includegraphics[height=5.2cm,clip]{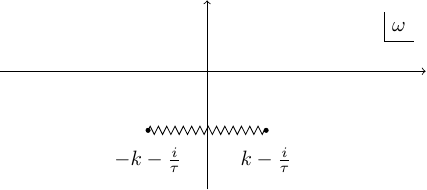}
		\caption{Illustration of branch cut in complex $\o$ plane. }
		\label{fig_cut}
	\end{center}
\end{figure}

To gain further analytic insights, we take the massless limit. We stress that the massless limit is only taken on the level of phase space integration without locking spin and momentum. The spin remains an independent field. We shall comment on the relaxation of massless condition shortly. In this limit, we can write the integral as
\begin{align}
&\int_p\d\widetilde{{\cal A}}^x=\int \frac{dp}{(2\p)^3} \big[-\frac{i\p p}{4k^3\t^2}\(-2k\t(i+\t\o)+\(k^2\t^2-(i+\t\o)^2\)\ln\frac{1+i\t(k-\o)}{1-i\t(k+\o)}\)\widetilde{\pd_3 F_{02}}f_V'\nonumber\\
&-\frac{i\p p}{2k^2\t}\(2k\t+(i+\t\o)\ln\frac{1+i\t(k-\o)}{1-i\t(k+\o)}\)\widetilde{\pd_0 F_{02}}f_V'-\frac{i\p p}{2k^2\t}k\t\ln\frac{1+i\t(k-\o)}{1-i\t(k+\o)}\widetilde{\pd_3 F_{02}}f_V'\nonumber\\
&+\frac{i\p}{4k^3\t^2}\(2k\t(i+\t\o)+(k^2\t^2+(i+\t\o)^2)\ln\frac{1+i\t(k-\o)}{1-i\t(k+\o)}\)\widetilde{\pd_3 F_{02}}f_V
\big],
\end{align}
where the overhead tilde indicates the corresponding quantity being Fourier transformed.
Using integration by part, we arrive at the following results after partial cancellation
\begin{align}
&\widetilde{J}_5^x=4\int_p\d\widetilde{{\cal A}}^x=\frac{iT \ln(1+e^{\m/T})}{(2\p)^2k^2\t}
\big[\(2k\t\ln\frac{1+i\t(k-\o)}{1-i\t(k+\o)}\)\widetilde{\pd_3 F_{02}}
\no
&+\(2k\t+(i+\t\o)\ln\frac{1+i\t(k-\o)}{1-i\t(k+\o)}\)\widetilde{\pd_0 F_{02}}\big].
\end{align}
We can reinstate rotational invariance by the replacement $\widetilde{\pd_3 F_{02}}\to-\e^{ijk}ik_j \tilde{E}_{k}$ and $\widetilde{\pd_0 F_{02}}\to\frac{\o}{k}\e^{ijk}ik_j \tilde{E}_{k}$ to have
\begin{align}\label{S_E}
&\widetilde{J}_5^i= \frac{iT \ln(1+e^{\m/T})}{(2\p)^2k^2\t}
\big[-\(2k\t\ln\frac{1+i\t(k-\o)}{1-i\t(k+\o)}\)\e^{ijk}ik_j \tilde{E}_{k}\no
&+\(2k\t+(i+\t\o)\ln\frac{1+i\t(k-\o)}{1-i\t(k+\o)}\)\frac{\o}{k}\e^{ijk}ik_j \tilde{E}_{k}\big].
\end{align}
\eqref{S_E} might not be illuminating. Let us expand in small $k$ as
\begin{align}\label{k_expand}
\widetilde{J}_5^i(\o,k\to 0)=\frac{T\ln(1+e^{\m/T})}{6\p^2}\frac{i(6i\t+5\t^2\o)}{(i+\t\o)^2}\e^{ijk}ik_j \tilde{E}_k+O(k^2).
\end{align}
Inverse Fourier transform gives
\begin{align}\label{local}
J_5^i(t,x)=\frac{T\ln(1+e^{\m/T})}{6\p^2}\int dt'\th(t')\(\frac{t'}{\t}+5\)e^{-t'/\t}\e^{ijk}\pd_j E_{k}(t-t',x).
\end{align}
We can further use Bianchi identity
\begin{align}\label{Bianchi}
\e^{ijk}\pd_j E_k=-\pd_0 B_i,
\end{align}
to rewrite \eqref{local} as
\begin{align}\label{J_B}
J_5^i(t,x)=-\frac{T\ln(1+e^{\m/T})}{6\p^2}\int dt'\th(t')\(\frac{t'}{\t}+5\)e^{-t/\t}\pd_0 B_i(t-t',x).
\end{align}
\eqref{J_B} is contribution from particle only. The anti-particle contribution is readily obtained by $\m\to-\m$ of \eqref{J_B} with an overall minus sign. The two contributions combined give
\begin{align}\label{J_B_sum}
J_5^i(t,x)=-\frac{\m}{3\p^2}\int dt'\th(t')\(\frac{t'}{\t}+5\)e^{-t'/\t}\pd_0 B_i(t-t',x).
\end{align}
The interpretation of \eqref{J_B_sum} is clear: the transient mode corresponds to retarded response to variation of magnetic field. Recall that our convention corresponds to a negatively charged fermion, which is clear from \eqref{off-shell}, thus the sign in \eqref{J_B_sum} is indeed consistent with retarded nature. Of course, the local nature of the response is a consequence of our small $k$ expansion. The actual response is non-local, which can be seen from the complicated $k$-dependence of \eqref{S_E}. The structure of the non-locality can be deduced from the logarithms: in $\o$ space, there is a branch cut along $(-k-\frac{i}{\t},k-\frac{i}{\t})$, as shown in Fig.~\ref{fig_cut}. We shall elaborate the origin of the branch cut soon.

Now we extend our discussion hydrodynamic source. 
We proceed to solve \eqref{transportH_RTA} for different sources. The explicit source corresponding to the shear is given by
\begin{align}
\frac{1}{2p_0}\e^{ijk}p_lp_k\s_{lj}\d(p^2-m^2)f_V',
\end{align}
for $\m=i$. Following the treatment in EM fields source, we point the flow along $y$-direction and inhomogeneity along $z$-direction without loss of generality. The source is nonvanishing for $i=x$ only, which leads to the following solution
\begin{align}
\widetilde{J}_5^x=4\int_p\frac{1}{4p_0}(p_3^2-p_2^2)\widetilde{\pd_3u_2}\d(p^2-m^2)f_V'.
\end{align}
We again take the massless limit and perform the angular integration to obtain
\begin{align}
\widetilde{J}_5^x=\int dp \frac{ip^2}{16\p^2k^3\t^3}\big[\(k^2\t^2-3(i+\t\o)^2\)\ln\frac{1+i\t(k-\o)}{1-i\t(k+\o)}-6k\t(i+\t\o)\big]\widetilde{\pd_3u_2}f_V'.
\end{align}
The $p$-integral for particle contribution is easily evaluated using
\begin{align}
\int dp \frac{p^2}{\p^2}f_V'=\frac{2T^2}{\p^2}Li_2(-e^{\m/T})\equiv-\c_+,
\end{align}
with $Li$ being the polylogarithm function.
Reinstating rotational invariance, we obtain
\begin{align}\label{S_sigma}
\widetilde{J}_5^i=- \frac{i\c_+}{8k^3\t^3}\big[\(k^2\t^2-3(i+\t\o)^2\)\ln\frac{1+i\t(k-\o)}{1-i\t(k+\o)}-6k\t(i+\t\o)\big]\e^{ijk}\hat{k}_j\hat{k}_l\tilde{\s}_{lk}.
\end{align}
Like in the EM fields case, the response is non-local from the logarithms in \eqref{S_sigma}. The location of the branch cut is at $\o\in(-k-\frac{i}{\t},k-\frac{i}{\t})$, which is the same as the EM field case.
Expansion in small $k$ gives
\begin{align}\label{S_shear_k0}
\widetilde{J}_5^i(\o,k\to0)=- \frac{i\c_+\t^2}{15(i+\t\o)^3}k^2\e^{ijk}\hat{k}_j\hat{k}_l\tilde{\s}_{lk}.
\end{align}
The anti-particle contribution is simply obtained with the replacement $\m\to-\m$. Note that there is no overall minus sign as spin-shear coupling is charge independent. We can combine the two contributions using the identity $Li_2(-x)+Li_2(-x^{-1})=\frac{\p^2}{6}+\frac{1}{2}\ln^2x$ to obtain
\begin{align}\label{S_shear_total}
\widetilde{J}_5^i(\o,k\to0)=- \frac{2i\c\t^2}{15(i+\t\o)^3}k^2\e^{ijk}\hat{k}_j\hat{k}_l\tilde{\s}_{lk},
\end{align}
where $\c=\frac{T^2}{6}+\frac{\m^2}{2\p^2}$. Using the definition of the shear tensor \eqref{shear_def} and vorticity $\O^i=\frac{1}{2}\e^{ijk}\pd_j u^k$, we can reexpress \eqref{S_shear_total} as spin response to inhomogeneity of vorticity
\begin{align}\label{S_nabla_vorticity_FF}
\widetilde{J}_5^i(\o,k\to0)=\frac{2i\c\t^2}{15(i+\t\o)^3}k^2\tilde{\O}^i.
\end{align}
We can transform back to coordinate space as
%
\begin{align}\label{S_nabla_vorticity}
{J}_5^i(t,x)&=-\frac{\c\t}{15}\int dt'\th(t')\(\frac{t'}{\t}\)^2e^{-t'/\t}\nabla^2\O^i(t-t',x).
\end{align}

We will not repeat the procedure for other hydrodynamic sources. We simply present results below. For acceleration source, we obtain
\begin{align}\label{S_acc}
\widetilde{J}_5^i(\o,k)=\frac{T^2\,Li_2(-e^{\m/T})}{16\p^2k^2\t^2}\(-2ik\t+(1-i\t\o)\ln\frac{1+i\t(k-\o)}{1-i\t(k+\o)}\)\e^{ijk}\hat{k}_j\widetilde{\pd_0u_k}.
\end{align}
In the small $k$ limit, we have
\begin{align}
\widetilde{J}_5^i(\o,k\to 0)=\frac{iT^2\t\,Li_2(-e^{\m/T})}{24\p^2(i+\t\o)^2}\e^{ijk}k_j\widetilde{\pd_0u_k}.
\end{align}
Adding a contribution from anti-particles and using definition of vorticity as in the shear case, we obtain
\begin{align}\label{S_time_vorticity_FF}
\widetilde{J}_5^i(\o,k\to 0)=-\frac{\c\,\t}{12(i+\t\o)^2}\widetilde{\pd_0\O^i}.
\end{align}
Inverse Fourier transform of it gives
%
\begin{align}\label{S_acc_total}
J_5^i(t,x)&=\frac{\c}{12}\int dt'\th(t')\frac{t'}{\t}e^{-t'/\t}\pd_0\O^i(t-t',x).
\end{align}
Clearly \eqref{S_acc_total} gives the retarded response to temporal variation of vorticity, which is similar to \eqref{J_B_sum}. 

The cases for temperature gradient and chemical potential gradient are trivial: the responses to these sources do not survive after the angular integration. This is because we cannot construct a pseudovector out of direction of spatial inhomogeneity. While in the cases of shear and acceleration, the fluid velocity provides a second necessary direction to saturate the indices.

We close this section by making the following observation: although we started out looking for spin responses to all possible dissipative sources, we have found all of them reducible to temporal and spatial variations of magnetic field and vorticity, see \eqref{J_B_sum}, \eqref{S_nabla_vorticity} and \eqref{S_acc_total}. These responses are higher order in gradient and are expected to modify spin hydrodynamics beyond lowest order.

\section{Implications for spin hydrodynamics}\label{sec_spinhydro}

The branch cut at $\o\in(-k-\frac{i}{\t},k-\frac{i}{\t})$ is one of the main results obtained in this work. It is present in responses to all sources. Since our results have been obtained in the massless limit, one may wonder about the validity of this conclusion beyond the massless limit. Below we argue that the presence of branch cut holds for massive case as well. Let us understand the origin of the branch cut from the schematic form \eqref{S_schematic}. There is pole in the response at $p_0/\t+i{\bf k}\cdot{\bf p}-i\o p_0=0$, which simplifies to $1/\t+i k\cos\th-i\o =0$ in the massless limit, with $\th$ being the angle between ${\bf k}$ and ${\bf p}$. The angular integration covers the range $\th\in(-1,1)$, giving rise to infinite many poles stretching along $\o\in(-k-\frac{i}{\t},k-\frac{i}{\t})$, which is equivalent to the branch cut. For massive case, the poles are located at
\begin{align}
\frac{1}{\t}+ik\frac{p}{p_0}\cos\th-i\o=0,
\end{align}
with $p_0=\sqrt{p^2+m^2}$. Angular integration gives rise to infinite many poles stretching along $\o\in(-k\frac{p}{p_0}-\frac{i}{\t},k\frac{p}{p_0}-\frac{i}{\t})$. We still have $p$-integration. The limit $p\to\infty$ corresponds to $\frac{p}{p_0}\to 1$. It follows that the location of the branch cut is unchanged, but the residues at each poles differ from the massless case. We close the argument by the remark that branch cut is generic in theories with quasi-particles. Examples of massless and massive cases can be found in \cite{Bellac:2011kqa} and \cite{Petitgirard:1991mf} respectively.

Now let us discuss the implication of our results. What we have considered are transient spin modes to EM and hydrodynamic sources without backreaction to the sources, i.e. in the probe limit. We shall make comparison with the results from spin hydrodynamics. For the canonical choice of spin tensor, the spin modes mix with the hydrodynamic modes from equation (4.3) of \cite{Hongo:2021ona} quoted below
\begin{align}\label{spinhydro}
&\pd_0\d\e+\pd_i\d\p^i=0,\no
&\pd_0\d\p_i+c_s^2\pd_i\d\e-\g_{\parallel}\pd_i\pd^j\d\p_j-(\g_\perp+\g_s)(\d_i^j\boldsymbol{\nabla}^2-\pd_i\pd^j)\d\p_j+\frac{1}{2}\G_s\varepsilon_{0ijk}\pd^j\d\s^k=0,\no
&\pd_0\d\s_i+\G_s\d\s_i+2\g_s\varepsilon_{0ijk}\pd^j\d\p^k=0.
\end{align}
$\d\e$ and $\d\p^i$ correspond to fluctuation of energy density and momentum density respectively, with the latter comes entirely from fluctuation of fluid velocity $\d\p^i=(\e_0+p_0)\d u^i$. $\e_0+p_0=h_0$ is the equilibrium enthalpy density. $\d\s^i$ corresponds to fluctuation of spin density, which is $\d J_5^i$ in our notation. The first equation is just conservation of energy. The second equation is conservation of momentum, subject to possible correction from curl of spin density. The third equation of \eqref{spinhydro} encodes the spin response to orbital angular momentum. It is easy to see that the mixing between hydrodynamic modes and spin modes occur between transverse momentum density and transverse spin density only (transverse refers to components perpendicular to momentum), with the following modification to the dispersion of the transverse spin modes
\begin{align}\label{spin_disp}
\o=-i\G_s\Rightarrow\o=-i\G_s-i\g_s k^2,
\end{align}
while the diffusive dispersion of transverse momentum density receives an additive correction in the diffusion constant
\begin{align}
\o=-i\g_\perp k^2\Rightarrow\o=-i(\g_\perp+\g_s) k^2.
\end{align}
In fact, we should be a little more careful with the mixing, which actually occurs between spin and combination of shear and vorticity. While the diffusive dispersion applies to the shear mode, the spin responds to vorticity only, which is clear from the third equation in \eqref{spinhydro}. It is the plane wave form used that includes both shear and vorticity as
\begin{align}
\pd_i u_j=\frac{1}{2}\(\pd_i u_j+\pd_j u_i\)+\frac{1}{2}\(\pd_i u_j-\pd_j u_i\),
\end{align}
with the symmetric and anti-symmetric terms corresponding to shear and vorticity contributions respectively\footnote{In principle, vorticity can have different dispersion from the shear. In particular a static vorticity can exist \cite{Becattini:2013fla} as well}. The same plane wave form is assumed in our analysis in the previous section, which allows us to rewrite spin response to shear tensor into counterpart to vorticity. With the spin responses to temporal and spatial variations of vorticity, we may ask how the last equation of \eqref{spinhydro} is modified.

In the static limit, we easily find from the balance of the last two terms that
\begin{align}
J_5^i=\frac{4\g_sh_0}{\G_s}\O^i=\c_s\O^i,
\end{align} 
using $\G_s=\frac{2\h_s}{\c_s}$ and $\g_s=\frac{\h_s}{2h_0}$ from \cite{Hongo:2021ona}. $\c_s$ is susceptibility of spin density, to be identified with $\c$ in our notation. Beyond the static limit, we may rewrite the response as
\begin{align}
J_5^i=-\frac{4\g_sh_0}{\G_s-i\o}\O^i=\frac{1}{1-i\o/\G_s}\c\O^i.
\end{align}
We see that $1/\G_s$ is the relaxation time, which is $\t$ in our notation. \eqref{S_nabla_vorticity_FF} and \eqref{S_time_vorticity_FF} suggest the following modification to the response
\begin{align}
\tilde{J}_5^i=\(\frac{1}{1-i\o\t}+\frac{-i\o\t}{12(1-i\o\t)^2}-\frac{2k^2}{15(1-i\o\t)^3}\)\c\tilde{\O}^i.
\end{align}
If we wish to convert the response to a local spin transport equation, we need to multiply both sides by $(1/\t-i\o)^3$ and perform inverse Fourier transform to obtain
\begin{align}\label{spin_transport}
(\pd_0+\frac{1}{\t})^3J_5^i=\(\frac{1}{\t}(\pd_0+\frac{1}{\t})^2+\frac{1}{12\t}(\pd_0+\frac{1}{\t})\pd_0+\frac{2}{15}\nabla\)\c\O^i
\end{align}
A striking feature is that the high order terms in gradients change the original first order equation into a third order one. This is in fact a consequence of non-local nature of the response. We can imagine if extending the analysis by including more terms in the expansion in $k$, we would end up with a transport equation for spin including more time derivatives. The observation suggests possible breakdown of expansion in $k$ in spin hydrodynamics beyond the lowest order for microscopic theories with quasi-particles. Nevertheless it is still meaningful to consider the limit $k=0$: in this case the branch cut shrinks to a point, the resulting transport equation is of second order as can be seen in \eqref{spin_transport} by dropping one overall $(\pd_0+\frac{1}{\t})$. The jump in order in the time derivatives can be understood from \eqref{S_acc_total}: since the spin also responds to temporal variation of vorticity, an extra condition specifying the temporal derivative of the vorticity is needed to fully determine the dynamics of the spin.

\section{Summary}\label{sec_sum}

We studied the dynamics of spin modes in the presence of dissipative sources including electric field and hydrodynamic gradients by solving the axial kinetic equation under the relaxation time approximation. We found transient spin modes in response to electric field with spacetime inhomogeneity, fluid acceleration and shear. To lowest order in spatial inhomogeneity $k$, we found the responses to electric field and fluid acceleration can be interpreted as retarded responses to temporal variations of magnetic field and fluid vorticity respectively. The response to shear arises at $O(k^2)$, which can be reduced to retarded response to spatial variation of fluid vorticity. Beyond the lowest order, we found the dispersions corresponding responses to all three sources contains the same branch cut indicating non-local nature of the responses. We argued that the non-locality is a consequence of the quasi-particle picture underlying the kinetic description.

We also discussed the implications of the transient spin modes to spin hydrodynamics. We pointed out that the known mixing between spin modes and shear modes actually involves both shear and vorticity modes. We reanalyzed the mixing between spin mode and vorticity mode alone taking into account spin responses to temporal and spatial variations of vorticity. It is found that the corrections turn the original first order spin transport equation into a third one if we aim at a local equation. In the homogeneous limit $k=0$, when the branch cut shrinks to a point, the resulting transport equation is of second order. The change in order of the equation is a consequence of non-local nature of the spin response to vorticity, suggesting possible breakdown of gradient expansion in spin hydrodynamics for microscopic theories with quasi-particles.

\section*{Acknowledgments}

We are grateful to Xu-Guang Huang for useful discussions. This work is in part supported by NSFC under Grant Nos 12075328, 11735007.

\appendix

\section{Source terms to ${\cal A}^\m$ with electric field}

In the appendix, we provide derivation of source terms in case of electric field. For convenience, we reproduce the corresponding kinetic equation
\begin{align}\label{AKT_E}
p\cdot\D \d\cA^\m+\frac{p\cdot u}{\t}\d\cA^\m=\frac{1}{2}\e^{\m\n\r\s}\pd_\s F_{\l\n}\pd_p^\l \cV_\r.
\end{align}
It can be rewritten as
\begin{align}
p\cdot\pd \d\cA^\m+s^\m_{\text{LHS}}=-p\cdot u\frac{\d\cA^\m}{\t}+s^\m_{\text{RHS}},
\end{align}
with
\begin{align}
&s^\m_{\text{RHS}}=\frac{1}{2}\e^{\m\n\r\s}(\pd_\s F_{\b\n})\d(p^2-m^2)f_V+\frac{1}{2}\e^{\m\n\r\s}(\pd_\s F_{\b\n})\big[2p_\r p^\b\d'(p^2-m^2)f_V+\no
&p_\r u^\b\d(p^2-m^2)f_V'\big],\no
&s^\m_{\text{LHS}}=p^\l\frac{1}{2}\e^{\m\n\r\s}p_\n \pd_\l F_{\r\s}\d'(p^2-m^2)f_V.
\end{align}
We have used the shorthand notation $f_V'=\frac{\pd f_V}{\pd(p\cdot u)}$.
The explicit forms of them are given by
\begin{align}\label{s_LH_explicit}
&s_{\text{RHS}}^0=\e^{ijk}\pd_kF_{0i}(-p_j)p_0\d'(p^2-m^2)f_V+\frac{1}{2}\e^{ijk}\pd_k F_{0i}(-p_j)\d(p^2-m^2) f_V',\no
&s_{\text{RHS}}^i=\e^{ijk}\pd_k F_{0j}\d(p^2-m^2) f_V+\e^{ilk}\pd_k F_{0l}p_0^2\d'(p^2-m^2)f_V+\e^{ijk}\pd_kF_{l0}p_jp_l\d'(p^2-m^2)f_V\no
&+\frac{1}{2}\e^{ijk}\pd_k F_{0j}p_0\d(p^2-m^2) f_V'+\e^{ijk}\pd_0 F_{0j}p_kp_0\d'(p^2-m^2)f_V+\frac{1}{2}\e^{ijk}\pd_0 F_{0j}p_k\d(p^2-m^2) f_V',\no
&s_{\text{LHS}}^0=0,\no
&s_{\text{LHS}}^i=-p_0p_j\e^{ijk}\pd_0F_{0k}\d'(p^2-m^2)f_V-p_lp_j\e^{ijk}\pd_l F_{0k}\d'(p^2-m^2)f_V.
\end{align}
There are three kinds of terms, which are proportional to $\d'(p^2-m^2)f_V$, $\d(p^2-m^2)f_V'$ and $\d(p^2-m^2)f_V$ respectively. We first collect terms of the first kind in $s^\m\equiv s^\m_{\text{RHS}}-s^\m_{\text{LHS}}$ as
\begin{align}\label{s_deltaprime}
&s^0=\e^{ijk}\pd_k F_{0i}(-p_j)p_0\d'(p^2-m^2)f_V,\no
&s^i=\e^{ilk}\pd_k F_{0l}p_0^2\d'(p^2-m^2)f_V+\e^{ijk}\pd_k F_{l0}p_jp_l\d'(p^2-m^2)f_V+p_lp_j\e^{ijk}\pd_l F_{0k}\d'(p^2-m^2)f_V.
\end{align}
Using the Schouten identity $p_l\e^{ijk}-p_i\e^{jkl}+p_j\e^{kli}-p_k\e^{lij}=0$, we can simplify $s^i$ in \eqref{s_deltaprime} as
\begin{align}\label{si_deltaprime}
&s^i=\e^{ilk}\pd_k F_{0l}p_0^2\d'(p^2-m^2)f_V+p_ip_j\e^{jkl}\pd_l F_{0k}\d'(p^2-m^2)f_V-{\bf p}^2\e^{ikl}\pd_l F_{0k}\d'(p^2-m^2)f_V\no
&=p_ip_j\e^{jkl}\pd_l F_{0k}\d'(p^2-m^2)f_V+m^2\e^{ilk}\pd_k F_{0l}\d'(p^2-m^2)f_V-\e^{ikl}\pd_k F_{0l}\d(p^2-m^2)f_V.
\end{align}
The last term cancels the only term of the kind $\d(p^2-m^2)f_V$ in \eqref{s_LH_explicit}. Adding the remaining terms of the kind $\d(p^2-m^2)f_V'$ to \eqref{si_deltaprime} and \eqref{s_deltaprime}, we obtain
\begin{align}\label{s_total}
&s^0=-p_0p_j\e^{ijk}\pd_k F_{0i}\d'(p^2-m^2)f_V+\frac{1}{2}\e^{ijk}\pd_k F_{0i}(-p_j)\d(p^2-m^2)f_V'\no
&s^i=p_ip_j\e^{jkl}\pd_l F_{0k}\d'(p^2-m^2)f_V+m^2\e^{ilk}\pd_k F_{0l}\d'(p^2-m^2)f_V+\frac{1}{2}\e^{ijk}\pd_k F_{0j}p_0\d(p^2-m^2)f_V'\no
&+\frac{1}{2}\e^{ijk}\pd_0 F_{0j}p_k\d(p^2-m^2)f_V'.
\end{align}

\bibliographystyle{unsrt}
\bibliography{transient.bib}

\end{document}